\documentclass[aps,pra,twocolumn,superscriptaddress,floatfix]{revtex4}

\usepackage{graphicx}
\usepackage{amssymb}
\usepackage{subfigure}
\usepackage{amsmath,array}
\usepackage{amsfonts}
\usepackage{bm}
\usepackage{color}
\usepackage[dvipsnames]{xcolor}
\usepackage{physics}
\usepackage{amsthm}
\usepackage{pgf,tikz}
\usetikzlibrary{arrows}
\usepackage{ulem}
\usepackage{verbatim}
\usepackage{hyperref}

\newcommand{\myeq}[1]{Eq.~(\ref{#1})}

\newcommand{\myfig}[1]{Fig.~\ref{#1}}
\theoremstyle{definition}
\newtheorem{definition}{Definition}[section]

\definecolor{mygreen}{RGB}{0,107,0}

\begin{document}

\title{Solving the One-dimensional Distance Geometry Problem by Optical Computing}

\author{S.B.~Hengeveld}
\email{simon.hengeveld@irisa.fr}
\affiliation{IRISA, University of Rennes 1, Rennes, France}
\author{N.~Rubiano~da~Silva}
\email{nara.rubiano@ufsc.br}
\affiliation{Departamento de F\'{i}sica, Universidade Federal de Santa Catarina, CEP 88040-900, Florian\'{o}plis, SC, Brazil}
\author{D. S. Gon\c{c}alves}
\email{douglas.goncalves@ufsc.br}
\affiliation{Department of Mathematics, CFM, Federal University of Santa Catarina, Florian\'{o}polis, SC, Brazil}
\author{P. H. Souto Ribeiro}
\email{p.h.s.ribeiro@ufsc.br}
\affiliation{Departamento de F\'{i}sica, Universidade Federal de Santa Catarina, CEP 88040-900, Florian\'{o}plis, SC, Brazil}
\author{A.~Mucherino}
\email{antonio.mucherino@irisa.fr}
\affiliation{IRISA, University of Rennes 1, Rennes, France}

\begin{abstract}
Distance geometry problem belongs to a class of hard problems in classical computation that can be understood in terms of a set of inputs processed according to a given transformation, and for which the number of possible outcomes grows exponentially with the number of inputs. It is conjectured that quantum computing schemes can solve problems belonging to this class in a time that grows only at a polynomial rate with the number of inputs. While quantum computers are still being developed, there are some classical optics computation approaches that can perform very well for specific tasks. Here, we present an optical computing approach for the distance geometry problem in one dimension and show that it is very promising in the classical computing regime.
\end{abstract} 

\maketitle

%%%%%%%%%%%%%%%%%%%%%%%%%%%%%%%%%%%%%%%%%%%%%%%%%%%%%%%%%%%%%%%%%%%%%%%%%%%%%%%%%%%%%%%%%%%%%%%%%
%%%%%%%%%%%%%%%%%%%%%%%%%%%%%%%%%%%%%%%%%%%%%%%%%%%%%%%%%%%%%%%%%%%%%%%%%%%%%%%%%%%%%%%%%%%%%%%%%
%%%%%%%%%%%%%%%%%%%%%%%%%%%%%%%%%%%%%%%%%%%%%%%%%%%%%%%%%%%%%%%%%%%%%%%%%%%%%%%%%%%%%%%%%%%%%%%%%

\section{Introduction}  \label{sec:intro}

The Distance Geometry Problem (DGP) asks whether a simple weighted undirected graph $G=(V,E,d)$ can be realized in the Euclidean space $\mathbb{R}^K$ so that the distance between two realized vertices corresponds to the real-valued weight $d(u,v)$, for all edges  $\{u,v\} \in E$. In other words, we look for vectors $x_1,\dots,x_n$ in $\mathbb{R}^K$, where $n=|V|$, such that $\| x_u - x_v \| = d(u,v), \forall \{u,v\} \in E$. 
The DGP is NP-hard \cite{saxe79}, and there exists a lot of scientific literature on this topic \cite{siam}. The general interest for this problem is related to graph theory, continuous and combinatorial optimization, as well as other problems in operational research. Apart from this, DGP has several interesting real-life  applications arising in different domains. 

One traditional and widely studied example is given by an application in structural biology, where the vertices of $G$ represent the atoms of a molecule, the distances are derived from experimental techniques such as Nuclear Magnetic Resonance, and the aim is to find the three-dimensional conformation of the molecule satisfying all distance constraints \cite{jcim19}. Another widely studied DGP application gives rise to the Sensor Network Localization Problem (SNLP), where the vertices of $G$ are sensors and the existence of an edge between two vertices indicates that they are in the range of communication, and the associated weight is an estimation of their distance \cite{Biswas06}.

In this work, we restrict our focus on DGPs arising from applications in dimension~1 \cite{edwco18}. An example of application in only one dimension is the Angular Synchronization Problem (ASP), where it is required to obtain an estimation (up to a constant additive phase) for a set of unknown angles by using the information about some measurements of their offsets, modulo $2\pi$ \cite{Singer11}. The problem of synchronizing the clocks in a distributed network \cite{giridhar06} is basically an ASP where the angles are replaced with time measures that are not periodic, making the problem actually correspond to a DGP in dimension~1.

We focus our attention on a class of BuildUp algorithms for the solution of DGPs that, in order to be used for finding DGP solutions, require the problem to satisfy some particular additional assumptions. We will introduce the notion of a {\it paradoxical} DGP instance in Section~\ref{sec:paradoxical}, which can possibly be solved by employing such BuildUp algorithms. However, this is done at an extremely high computational cost. This computation cost is a consequence of the NP-hardness of the general DGP, but it is because of the particular structure of the paradoxical instances that the complexity becomes extremely high. In practice, the complexity of these BuildUp algorithms tend to increase exponentially with the size of the DGP instance.

After the reformulation of the paradoxical DGP as a matrix-by-vector multiplication problem (this reformulation does not have any relevant impact on the total complexity; the matrix contains an exponentially growing number of rows), we propose the use of an optical scheme for performing this multiplication. In fact, the optical scheme can take into consideration several rows of the matrix {\it at the same time}, which are processed by a single light beam. Ideally, when the entire matrix can be processed in one unique operation by the optical scheme, the total complexity falls to~1, because all mathematical operations required for the matrix-by-vector multiplication are performed in parallel.

We point out that the practical realization of the proposed optic scheme cannot (at least with current technologies) achieve these idealistic performances on medium and large sized problem instances. However, it represents a powerful strategy for speeding up the computations required to solve the hardest instances (the paradoxical instances) of the DGP.

Optical computing schemes have recently been appearing in the scientific literature for attempting the solution of NP-hard problems in a processing time consistently faster than the one attainable by standard computers. In \cite{oltean2009} and \cite{xu2020}, for example, two optic schemes for the Subset Sum Problem (SSP) have been proposed; the SSP is an NP-hard problem which is quite ``close'' to the DGP as it is possible to reduce, via a polynomial transformation, the SSP to a unidimensional DGP \cite{coap12}. Another optic scheme for the well-known Traveling Salesman Problem (TSP) is proposed in \cite{haist2007}. Our work instead takes its basis from the optic scheme for the TSP initially proposed in \cite{shaked2007}, which is implemented using frequency combs and programmable transmissive devices. We present a spatial version of this approach.

The remaining of the paper is organized as follows. In Section~\ref{sec:paradoxical}, we will formally introduce the {\it paradoxical} DGP, which is a special class of DGPs where all instances have maximal exponential complexity (recall that the DGP is, in general, an NP-hard problem). In this same section, we will briefly discuss on BuildUp algorithms for the solution of DGPs, and we will implement one of such algorithms as a matrix-by-vector multiplication procedure. Our optical computing approach, based on the matrix-by-vector multiplication, will be presented in Section~\ref{sec:optic-scheme}. Finally, Section~\ref{sec:concl} will conclude the paper with a discussion and future works.

%%%%%%%%%%%%%%%%%%%%%%%%%%%%%%%%%%%%%%%%%%%%%%%%%%%%%%%%%%%%%%%%%%%%%%%%%%%%%%%%%%%%%%%%%%%%%%%%%
%%%%%%%%%%%%%%%%%%%%%%%%%%%%%%%%%%%%%%%%%%%%%%%%%%%%%%%%%%%%%%%%%%%%%%%%%%%%%%%%%%%%%%%%%%%%%%%%%
%%%%%%%%%%%%%%%%%%%%%%%%%%%%%%%%%%%%%%%%%%%%%%%%%%%%%%%%%%%%%%%%%%%%%%%%%%%%%%%%%%%%%%%%%%%%%%%%%

\section{Paradoxical distance geometry}  \label{sec:paradoxical}

An instance of the DGP can be represented by a simple weighted undirected graph $G=(V,E,d)$, where the existence of an edge $\{u,v\} \in E$ between $u$ and $v \in V$ indicates that the distance between the two vertices is known \cite{siam}. The weight $d(u,v)$ associated with the edge $\{u,v\}$ is the numerical value of the distance, which it is supposed to be {\it exact}, i.e.~very precise, in this work.

BuildUp algorithms for the DGP are based on the idea to construct DGP solutions by realizing one vertex per time, and in a predefined order \cite{Dong03}. Some of the algorithms based on this approach are able to obtain one unique realization for each vertex of the graph, while others are instead based on the generation of a search tree, where multiple solutions can be identified \cite{Carvalho08}. Several algorithms belonging to the latter subclass of BuildUp algorithms have recently been proposed in the scientific literature, because they are able to satisfy the assumptions given by real-life applications. One important member of this subclass is the Branch-and-Prune (BP) algorithm \cite{lln08}, that we will briefly sketch in Section~\ref{sub:bp} for the unidimensional case.

In this work, we focus our attention on a special class of DGPs in dimension~1 (that we will denote by DGP$_1$), where the graph $G$ satisfies the following additional properties.
\begin{definition}  \label{def:paradoxDGP1}
A simple weighted undirected graph $G=(V,E,d)$ represents a ``paradoxical'' instance of the DGP$_1$ if and only if $G$ satisfies the following assumptions:
\begin{enumerate}
%\item $G$ is connected;
\item there exists a vertex order on $G$ such that, for every vertex $k$ (different from the first vertex in the order), the edge $\{k-1,k\}$ belongs to $E$;
\item no other edge is included in $E$ except the edge $\{1,n\}$, providing the distance information from the first and the last vertex in the order ($n = |V|$).
\end{enumerate}
Since it is supposed that a vertex order exists on $G$, we use the rank $k$ for indicating the corresponding vertex $v$ in $V$.
\end{definition}

Notice that assumption~1 in the definition implies that $G$ is a connected graph. Moreover, since we focus on the unidimensional case, we can also remark that assumption~2 implies that $G$ is a cycle. 

We point out that the adjective {\it paradoxical} comes from the fact that, from the one hand, the presence of the last distance $d(1,n)$ %allows us to select only 2~solutions (symmetric w.r.t.~the position of the first vertex of $G$) out of an exponential number of solutions \cite{powerOf2}; 
implies that the problem has only 2~solutions (symmetric w.r.t.~the position of the first vertex of $G$) out of an exponential number of candidate realizations \cite{powerOf2} (see~Subsection~\ref{sub:bp}); 
on the other hand, the fact that the distance $d(1,n)$ is related to the very first and the very last vertex of $G$ implies that we can actually exploit it only when potential positions for the last vertex are constructed \cite{chapterDGA}. Therefore, when the BP algorithm is used to solve paradoxical DGP instances, the number of possible positions for a vertex keeps growing exponentially as we step from one vertex to the next in the order, until the very last vertex is reached. It is only here, when the last vertex is reached, that we can finally exploit the distance $d(1,n)$ and select the only two solutions that we expected to obtain.

\subsection{The Branch-and-Prune algorithm}  \label{sub:bp}

When a DGP$_1$ instance satisfies %all properties given above, 
the assumptions in Definition~\ref{def:paradoxDGP1}, 
we can run the following procedure for finding its solutions. First of all, we can initialize the position of the vertex $1$ in the origin of the %Cartesian system associated to our $1$-dimensional Euclidean space. 
real line. 
Let $x_1$ be this position, set to $0$. Since by hypothesis the distance $d(k-1,k)$ is known for every $k > 1$, we can compute the two possible positions for all other subsequent vertices $k$ by applying the formula:  
\begin{equation}  \label{eq:branching}
x_k = x_{k-1} + s_k d_{k-1,k}, \quad \forall k = 2,\dots, n ,
\end{equation}
where the value of $s_k$ is $+1$ when computing the first position, and it is instead set to $-1$ in the second computed position for $x_k$. Since there are potentially two distinct positions for $x_k$ for {\it every} previously computed position for $x_{k-1}$, the set of all vertex positions admits the structure of a tree, where layers contain the various positions for the same vertex $k$, and a path from the root node (the vertex $1$) to any of the leaf nodes represents a %solution 
candidate realization to our instance of the DGP$_1$ (a more detailed description is given in Appendix~A). For a paradoxical DGP$_1$ instance, among all candidate realizations, only those verifying $|x_1 - x_n| = d(1,n)$ are solutions to the problem.

This procedure is at the basis of an important algorithm that was proposed in the context of the DGP in 2008 in \cite{lln08}, the Branch-and-Prune (BP) algorithm, and several times revised and adapted to different settings since then. The {\it branching} part in the name is related to procedure that is able to compute $2 \times p$ new candidate vertex positions for the vertex $k$, when $p$ positions are already available for the vertex $k-1$. In our specific case, the branching procedure reduces to applying \myeq{eq:branching} with the two possible choices for $s_k$.

The {\it pruning} part consists in exploiting additional distance information, that was not used during branching, with the aim of verifying the feasibility of computed vertex positions w.r.t.~the entire distance information. If some of these additional distances are not satisfied, then the current position, and hence the entire tree branch having as a root this position, can be pruned away from the tree. In the case of paradoxical DGP instances, these additional distances are actually absent until the layer $n$ of the tree is reached. It is only there where two distances (and not only one) are available to find out the potential positions for the vertex $n$, the last in the vertex order. The first distance, say $d(n-1,n)$, can be used for branching, by applying \myeq{eq:branching} as in the previous steps. Then, the second distance, say $d(1,n)$, can be used for pruning purposes: all positions that were generated during the branching phase such that $|x_1 - x_n| \ne d(1,n)$ can actually be removed from the tree.

In our particular settings, we can slightly modify this algorithm in order to have an uniform treatment of the distance information over the tree, without explicitly distinguishing between distances used for branching or for pruning purposes. This uniformity will help us then to construct the optic scheme, because it will avoid us to implement any conditional process. If the DGP$_1$ instance at hand is a paradoxical instance, then we can add a fictive vertex, name it ``$n+1$'', and add it to the original graph $G$. Then, remove the original edge $\{1,n\}$ and add the new edge $\{n,n+1\}$, where the weight $d(n,n+1)$ corresponds to the distance value $d(1,n)$. Now, it is straightforward to verify that \myeq{eq:branching} can be applied over all tree layers, in the given vertex ordering from $2$ to $n+1$, and that the the solutions to this DGP are all realizations for which
\begin{displaymath}
x_1 = x_{n+1} . 
\end{displaymath}
Notice that, if there exists no realization satisfying this identity, then the DGP instance at hand is infeasible.

\subsection{A matrix-by-vector reformulation}  \label{sub:reformulation}

Let $G=(V,E,d)$ be a simple weighted undirected graph representing a paradoxical instance of the DGP$_1$ (see Section~\ref{sec:intro}). As proposed in \cite{shaked2007} for the optic computing approach to TSP, we reformulate our special class of the DGP$_1$ so that the computation core reduces to a matrix-by-vector multiplication. We point out that the construction of the matrix to perform this multiplication can be done by defining and applying special matrix operators a polynomial number of times \cite{shaked2007}. However, when the matrix needs to be explicitly constructed element by element, the overall complexity grows exponentially: this is the reason why we will not propose in this work any construction procedure making use of matrix operators, but directly a procedure for an explicit element by element construction.

The basic idea behind the reformulation of our paradoxical DGP$_1$ as a matrix-by-vector multiplication consists in representing a 
%solution to the problem 
candidate realization 
as a boolean row $n$-vector, where the values of the elements are either $-1$ or $+1$. For example, the vector
\begin{displaymath}
(-1, +1, +1, \dots, -1, -1),
\end{displaymath}
representing a realization, 
when multiplied by the column vector $\mathbf{y}$, defined as
\begin{displaymath}
\mathbf{y} = \left[
 \begin{array}{c}
 d_{12} \\
 d_{23} \\
 \vdots \\
 d_{n-1,n} \\
 d_{1n}
 \end{array}
 \right] ,
\end{displaymath}
provides the position of the fictive vertex ``$n+1$''. As indicated above, if $x_1=0$, then this realization is feasible if and only if %the position for the fictive vertex is $0$ as well.
$x_{n+1}=0$ as well. 

In order to take into consideration all possible realizations at once, we define a matrix $\mathbf{M}$ which is capable of holding all possible boolean vectors enconding these realizations. Every row of this matrix corresponds to one realization: since the total number of potential realizations is $2^{n-1}$, $\mathbf{M}$ has, in total, $2^n$ rows (because of the addition of a fictive vertex); moreover, since the number of vertices is $n$, $\mathbf{M}$ has $n$ columns.

While the construction of the vector $\mathbf{y}$ is trivial (it simply consists of the list of available distances, in the order implied by the vertex order), we have derived a simple formula for an efficient construction of the matrix $\mathbf{M}$. To introduce this formula in general for every problem size $n > 0$, let us first of all consider the simple case where only one vertex is contained in the graph $G$, so that only the last distance $d_{1n}$ is theoretically present (notice that this is a special instance admitting no solutions unless $d_{1n} = 0$). In this case, we have
\begin{equation}  \label{equ:caseOneVertex}
\mathbf{M} = \left(
\begin{array}{r}
-1 \\
 1 \\
\end{array}
\right) , \qquad
\mathbf{y} = ( d_{1n} ) ,
\end{equation}
and there are only two potential solutions, represented by the elements of the vector  $\mathbf{r}^T = (-d_{11},d_{11})$. 
% \begin{displaymath}
% \mathbf{s} = \left(
% \begin{array}{r}
% -d_{11} \\
%  d_{11} \\ 
% \end{array}
% \right) .
% \end{displaymath}
As remarked above, $\mathbf{M}$ is a $2^n \times n$ matrix in general. In the first (and unique) column for the matrix $\mathbf{M}$ in \myeq{equ:caseOneVertex}, we can remark that the element equal to $-1$ is the one with the row index $i$ that is odd, whereas the element equal to $1$ has the even row index (we suppose that the first index is~1; the formulae can be simply adapted to the case where the first index would instead be~0). Let us add now a second vertex to our instance, in a way that it still satisfies the properties of a paradoxical DGP$_1$ instance. The matrix and vector to be multiplied are now:
\begin{displaymath}
\mathbf{M} = \left(
\begin{array}{rr}
-1 & -1 \\
 1 & -1 \\
-1 &  1 \\
 1 &  1 \\
\end{array}
\right) , \qquad
\mathbf{y} = \left(
\begin{array}{c}
d_{12} \\
d_{12} \\
\end{array}
\right),
\end{displaymath}
resulting in $\mathbf{r}^T = (-2d_{12},0,0,2d_{12})$.
Again, this is only a theoretical instance, which always admits two solutions (the ones corresponding to the rows $(-1,1)$ and $(1,-1)$, for every possible value for the distance $d_{12}$). Even if only theoretical, this instance allows us to have a clearer idea on the general formula for the construction of the matrix $\mathbf{M}$. It is easy to verify that the same rule identified above is still true for the new rows of the matrix, as far as the first column is concerned. This rule is equivalent to saying that the element value is $-1$ if and only if $(i-1)\bmod 2$ is zero. For the second column, a similar rule can be identified: the element is $-1$ if and only if $(i-1)/2\bmod 2 = 0$, where $(i-1)/2$ is the integer division. Notice the fact that the row indices are divided by~2 in the rule concerning the second column. 

From these rules the following general formula can be derived:
\begin{displaymath}
\mathbf{M}_{ij} = \left\{
\begin{array}{rl}
-1 & {\rm if} \, (i-1) 2^{1-j} \bmod 2 = 0 , \\
 1 & {\rm otherwise} . \\
\end{array}
\right.
\end{displaymath}

We can remark that the definition of every element of the matrix does not depend on any of the others, so that every submatrix of $\mathbf{M}$ can be easily constructed. This property of our general formula has a very important, positive impact on the presented optimal scheme (see next section), because the resolution limitations of our optic devices may bring us to split up the calculations (and therefore the matrix $\textbf{M}$) in several parts. We can also remark that, in case it is only necessary to know whether the DGP$_1$ instance at hand is feasible or not (without explicit reconstruction of the solution if the instance is actually feasible), then the feasibility can simply be verified by checking whether the vector $\mathbf{r} = \mathbf{My}$ (resulting from the matrix-vector multiplication) %$\mathbf{M} \times \mathbf{y}$ 
has a null component.

%%%%%%%%%%%%%%%%%%%%%%%%%%%%%%%%%%%%%%%%%%%%%%%%%%%%%%%%%%%%%%%%%%%%%%%%%%%%%%%%%%%%%%%%%%%%%%%%%
%%%%%%%%%%%%%%%%%%%%%%%%%%%%%%%%%%%%%%%%%%%%%%%%%%%%%%%%%%%%%%%%%%%%%%%%%%%%%%%%%%%%%%%%%%%%%%%%%
%%%%%%%%%%%%%%%%%%%%%%%%%%%%%%%%%%%%%%%%%%%%%%%%%%%%%%%%%%%%%%%%%%%%%%%%%%%%%%%%%%%%%%%%%%%%%%%%%

\section{Optical computation scheme}  \label{sec:optic-scheme}

\begin{figure*}[!t]
\includegraphics[width=\textwidth]{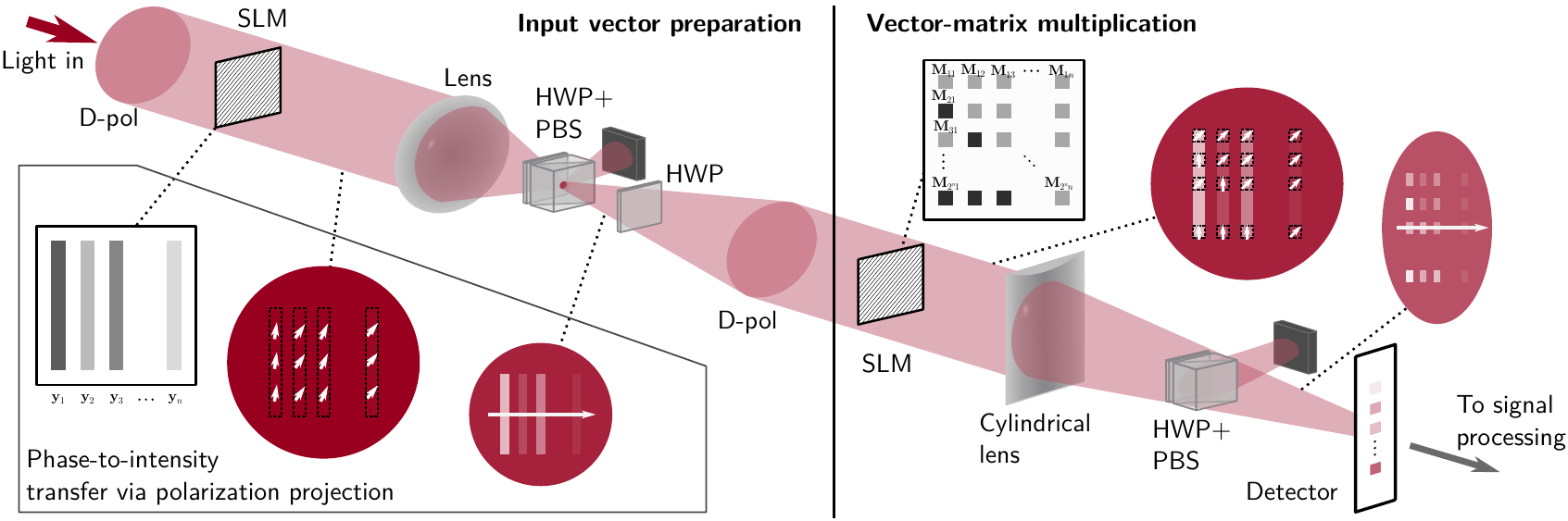}
\caption{Optical scheme to perform multiplication-accumulation (MAC) operations. An input vector is prepared using a first SLM; the vector is then multiplied by a matrix using a second SLM (which encodes the matrix elements) and a cylindrical lens. The resulting vector is detected using a CCD camera, or photodiodes. The insets show the gray level distribution on the SLMs, and the beam transverse profile at different stages (arrows indicate polarization, color level indicate intensity).}
\label{fig:macoperations}
\end{figure*}
%
%This is a scheme useful 
%only if we can trasform the DGP or the subset-sum problem (or any other you find interesting) 
We now present a possible optical scheme that performs matrix-vector multiplications, and is thus useful to solve a DGP$_1$ instance, according to the reformulation given in the previous section.
We developed this idea on the basis of the recent works presented in \cite{feldmann2021}.

We make use of light polarization and transverse spatial distribution of phases in a light beam to realize the required operations. Figure \ref{fig:macoperations} (top row) shows the preparation of the vector $\mathbf{y}$ (transposed to a line vector). A light beam polarized in the diagonal direction is sent to a spatial light modulator (SLM) that encodes the vector elements ($d_{i,i+1}$, with $i=1,\dots,n$) in a stripe-like phase mask. The numerical values in the vector are mapped onto gray levels of the SLM pixels, which in turn imprint phase shifts to the incoming light according to these levels. However, the SLM acts only in the horizontal polarization component, so that the phase applied by the SLM becomes a phase difference between vertical and horizontal polarization components. As a result, the polarization state is locally modified within the wavefront. The outgoing light beam passes through a half-wave plate (HWP) and is filtered by a polarizing beam splitter (PBS), which selects only the diagonal polarization component. This operation converts the phase modulation into an amplitude modulation. Therefore, the vector elements are encoded in the amplitude transverse spatial distribution of the beam in the image plane of a lens.

The combination of phase-only SLMs with before- and after-polarization projections is a method to convert the SLM-imprinted phase pattern into the amplitude distribution of the light beam \cite{paul2014}. One advantage of using the amplitude-related encoding instead of phase is that amplitude information can be straightforwardly measured through the intensity distribution and recorded as an image using, e.g., a CCD camera.

% Again, the scheme is based on a light beam and phase-only SLMs. Similar to the scheme described in \ref{sec:slmsequence}, in order to transform the phase maps into amplitude maps, the light must have diagonal polarization before impinging on the SLMs (acting only on the horizontal polarization), and the resulting beam must be projected in a proper polarization state (namely, not vertical).

The processing of a matrix-vector multiplication (MVM) is displayed in \myfig{fig:macoperations}. The input vector encoded in the light beam prepared by the first SLM is imaged onto the surface of a second SLM. Its polarization state is previously rotated to diagonal and the same strategy used in the first step is repeated. The second SLM is programmed with gray levels that encode the matrix elements. Specifically, only two gray levels are needed for solving a DGP$_1$ instance, since the elements of the matrix $\mathbf{M}$ may have only two values: -1 or 1. The exit beam will have a spatial modulation in polarization, corresponding to the multiplication of each vector element by each matrix row element. To complete the (MVM) operation, the elements of each row must be accumulated. This is implemented by a cylindrical lens focusing along the appropriated row direction. The polarization is filtered again keeping the horizontal component, and we obtain the resulting vector elements as an intensity map, which is subsequently recorded using a CCD camera or photodiodes.

The output signal is thus composed by $2^n$ ``pixels'', corresponding to each of the $2^n$ elements of the vector $\mathbf{r}$. We are particularly interested in knowing whether there are feasible solutions, which can be directly verified by the presence of a pixel with the intensity encoding a value 0, i.e., a null component of the vector $\mathbf{r}$, as discussed in the previous section.

\subsection{Connection between intensity measurements and the target operations}

We will now explicitly show how the intensity measurements performed using the proposed scheme relate to the target matrix-by-vector multiplication. For that, we first demonstrate the phase-to-intensity transfer via polarization projection, taking into account the effects of an SLM, an HWP and a PBS on an input field (see lower inset in \myfig{fig:macoperations}).

Let us suppose an input light field (before the SLM) as an ideal plane wave propagating along the $z$-direction, whose electric field complex amplitude is given by:
\begin{equation}
\label{eq:field1}
\vec{E}^{in} = E_0 \,\, \vec{\epsilon}\,\, \mbox{e}^{-i(\vec{k} \cdot \vec{r} - \omega t)} ,
\end{equation}

\noindent where $E_0$ is a constant, $\vec{\epsilon}$ is the polarization vector, $\vec{k}~=~k\hat{z}$ is the wave vector in vacuum, and $\omega$ is the optical frequency. Assuming a monochromatic field, the temporal phase $\omega t$ will only contribute to a global phase and will not play a role in our scheme; $\omega t$ can hence be omitted. We further assume that the SLM is located at the plane $\vec{r} = (x,y,0)$ and the polarization state is linear diagonal $\vec{\epsilon} = (\hat{x} + \hat{y})/\sqrt{2}$.

The input field is modified by the SLM, which imprints a phase modulation on the horizontal polarization component. The modulation is given by the M~$\times$~N matrix:
\begin{displaymath}
S = 2
\left[
  \begin{array}{cccccc}
     & \phi_{11} & \phi_{12} & \hdots & \phi_{1n} &  \\
     & \phi_{21} & \phi_{22} & \hdots & \phi_{2n} &  \\
     & \vdots & \vdots & \ddots & \vdots &  \\
     & \phi_{m1} & \phi_{m2} & \hdots & \phi_{mn} & \\ 
  \end{array}
\right],
\end{displaymath}
\noindent where $\phi_{mn}$ is the phase added to the horizontal component of the field impinging on cell $mn$, at which the field amplitude is $E_0/(MN)$. Therefore, the field just after the SLM:
\begin{equation}
\label{eq:field2}
\vec{E}_{mn}^{SLM} = \frac{E_0}{MN \sqrt{2}} \,\, (\hat{x} + \mbox{e}^{-i2\phi_{nm}} \hat{y}),
\end{equation}
\noindent shows a spatial modulation on the polarization. In order to transfer such modulation to the intensity profile, a proper polarization projection must be realized. Specifically, we employ a HWP (tilting the polarization in $-45^\cdot$, i.e., D$\rightarrow$H and A$\rightarrow$V) and a PBS (filtering only the H component) to project the diagonal component of the field in \myeq{eq:field2}, resulting in \cite{paul2014}:
\begin{equation}
\label{eq:field3}
\vec{E}_{mn}^{out} = \frac{E_0}{2 MN } \,\, \cos(\phi_{mn})\hat{x}.
\end{equation}

Equation~\ref{eq:field3} clearly shows that the gray-level mask displayed on the SLM surface, $S$, is transferred to the transverse intensity distribution of the field. In other words, the combination of phase-only SLM, HWP and PBS works as an effective amplitude modulator.

Now, considering the whole proposed scheme (\myfig{fig:macoperations}), the operation just described is applied two times, sequentially: firstly for preparing the input vector, and then for multiplying it by a matrix. Specifically, a field given by \myeq{eq:field3}, with $\phi_{mn}$ relating to the first SLM, has its polarization tilted to the diagonal direction and then used as the input field for the second SLM, whose modulation we call $\alpha_{pq}$, so that the final field is given by:
\begin{equation}
\label{eq:field4}
\vec{E}_{mn}^{out} = \frac{E_0}{2\sqrt{2} MN } \,\, \cos(\phi_{mn}) \cos(\alpha_{pq}) \hat{x}. 
\end{equation}

We note that an imaging lens is used to carefully transfer the spatial structure of the beam from SLM1 to SLM2, ideally without any distortion, ensuring a proper overlap between the masks (both of size M~$\times$Ñ). The total amplitude modulation can be viewed in terms of the matrix $S_{amp}$, comprising the element-by-element multiplication:
\begin{widetext}
\begin{displaymath}
S_{amp} = 
\left[
  \begin{array}{cccccc}
     & \cos(\phi_{11})\cos(\alpha_{11}) & \cos(\phi_{12})\cos(\alpha_{12}) & \hdots & \cos(\phi_{1n})\cos(\alpha_{1q}) &  \\
     & \cos(\phi_{21})\cos(\alpha_{21}) & \cos(\phi_{22})\cos(\alpha_{22}) & \hdots & \cos(\phi_{2n})\cos(\alpha_{2q}) &  \\
     & \vdots & \vdots & \ddots & \vdots &  \\
     & \cos(\phi_{m1})\cos(\alpha_{p1}) & \cos(\phi_{m2})\cos(\alpha_{p2}) & \hdots & \cos(\phi_{mn})\cos(\alpha_{pq}) &  \\      
  \end{array}
\right].
\end{displaymath}
\end{widetext}

In the last step of the scheme, the beam is focused in the detection plane with a cylindrical lens, so that the lines of the matrix concentrate to a single cell, effectively performing the sum over the elements of each line. The field at each cell $p$ is given by:
\begin{equation}
\label{eq:field5}
\vec{E}_{p}^{DET} = \sum_q \frac{E_0}{2\sqrt{2}MN} \,\, \cos(\phi_{mn})_{pq}\cos(\alpha_{pq}) \hat{x}. 
\end{equation}

Apart from a constant multiplicative term, the transverse spatial distribution of the field is now represented by a column (it could be a line as well) vector $V^{DET}$:
\begin{widetext}
\begin{displaymath}
V^{DET} =
\left[
  \begin{array}{ccc}
     & \cos(\phi_{11})\cos(\alpha_{11}) +  \cos(\phi_{12})\cos(\alpha_{12}) + \hdots + \cos(\phi_{1n})\cos(\alpha_{1q}) &  \\
     & \cos(\phi_{21})\cos(\alpha_{21}) + \cos(\phi_{22})\cos(\alpha_{22}) + \hdots + \cos(\phi_{2n})\cos(\alpha_{2q}) &  \\
     & \vdots &  \\
     & \cos(\phi_{m1})\cos(\alpha_{p1}) + \cos(\phi_{m2})\cos(\alpha_{p2}) + \hdots + \cos(\phi_{mn})\cos(\alpha_{pq}) &  \\      
  \end{array}
\right].
\end{displaymath}
\end{widetext}

The result of the operation is given by the intensities that are measured by a CCD camera:
\begin{equation}
\label{eq:intensity}
{\cal I}_{p}^{DET} = \left| \vec{E}_{p}^{DET} \right|^2 
\equiv I_0 |V_{p}^{DET}|^2,
\end{equation}

% \begin{eqnarray}
% \label{eq:intensity}
% {\cal I}_{p}^{DET} &=& \left|\sum_p \frac{E_0}{2\sqrt{2}MN} \,\, \cos(\phi_{mn})_{pq}\cos(\alpha_{pq})\right|^2 , \\ \nonumber
% % &\equiv& I_0 \left|\sum_p \,\, \cos(\phi_{mn})_{pq}\cos(\alpha_{pq})\right|^2,\\ \nonumber
% &\equiv& I_0 |V^{DET}|^2,
% \end{eqnarray}

\noindent where $I_0 = \left|\frac{E_0}{2\sqrt{2}MN}\right|^2$. We can hence rewrite $V^{DET}$ in terms of the measured intensities:
\[
V^{DET} = 
\left[
  \begin{array}{ccc}
     & \sqrt{{\cal I}_{1}/I_0} &  \\
     & \sqrt{{\cal I}_{2}/I_0} &  \\
     & \vdots & \\
     & \sqrt{{\cal I}_{p}/I_0} &  \\      
  \end{array}
\right].
\]

Finally, the detected signal given by Eq. \ref{eq:intensity} can be identified with the multiplication of a vector $\mathbf{y}$ 
\[
\mathbf{y} = 
\left[
  \begin{array}{ccc}
     & \cos(\phi_{1}) &  \\
     & \cos(\phi_{2}) &  \\
     & \vdots &  \\
     & \cos(\phi_{m}) &   
  \end{array}
\right] 
\]
\vspace*{1cm}
by a matrix $\mathbf{M}$:
\vspace*{1cm}
\[
\mathbf{M} = 
\left[
  \begin{array}{cccccc}
     & \cos(\alpha_{11}) & \cos(\alpha_{12}) & \hdots & \cos(\alpha_{1q}) &  \\
     & \cos(\alpha_{21}) & \cos(\alpha_{22}) & \hdots & \cos(\alpha_{2q}) &  \\
     & \vdots & \vdots & \ddots & \vdots &  \\
     & \cos(\alpha_{p1}) & \cos(\alpha_{p2}) & \hdots & \cos(\alpha_{pq}) &  
  \end{array}
\right],
\]

\noindent 
by setting, for the first SLM, all elements of a column in $S$ to the same modulation: $\cos(\phi_{j1}) \equiv \cos(\phi_{1})$, $\cos(\phi_{j2}) \equiv \cos(\phi_{2})$, and so on. In particular, for solving a paradoxical DGP$_1$ instance, the elements of the matrix $\mathbf{M}$ should be mapped to the values $\{-1,1\}$, while the elements of the distances vector $\mathbf{y}$ must be mapped to the interval $[0,1] \subset [-1,1]$. This results in $V^{DET} = \mathbf{M} \mathbf{y}$:
\vspace*{1cm}
\begin{widetext}
\begin{displaymath}
V^{DET} = 
\left[
  \begin{array}{ccc}
     & \cos(\phi_{1})\cos(\alpha_{11}) + \cos(\phi_{2})\cos(\alpha_{12}) + \hdots + \cos(\phi_{m})\cos(\alpha_{1q}) &  \\
     & \cos(\phi_{1})\cos(\alpha_{21}) + \cos(\phi_{2})\cos(\alpha_{22}) + \hdots + \cos(\phi_{m})\cos(\alpha_{2q}) &  \\
     & \vdots &  \\
     & \cos(\phi_{1})\cos(\alpha_{p1}) + \cos(\phi_{2})\cos(\alpha_{p2}) + \hdots + \cos(\phi_{m})\cos(\alpha_{pq}) &  \\      
  \end{array}
\right].
\end{displaymath}
\end{widetext}

\subsection{Computating performance}

As a comparison, the achieved computing speed is 2~Tera-(MAC operations) per second in the experiments described in \cite{feldmann2021}. Using the full resolution of typical SLMs (i.e., one vector or matrix element per pixel), we could achieve 0.1~Tera-(MAC operations) per second. For that, we take the $1920 \times 1080$ resolution as the maximum matrix size, 1080 as the maximum number of input vectors that can be processed in parallel (in the same run of the processor), and the nominal frame rate of 60~Hz (the speed at which the pixels of the SLM can be changed to another value). Please note that we have not taken into account neither the speed of the light signal nor the electronic signal (of the detection part); the former, however, barely contributes to the processing time. 

%%%%%%%%%%%%%%%%%%%%%%%%%%%%%%%%%%%%%%%%%%%%%%%%%%%%%%%%%%%%%%%%%%%%%%%%%%%%%%%%%%%%%%%%%%%%%%%%%
%%%%%%%%%%%%%%%%%%%%%%%%%%%%%%%%%%%%%%%%%%%%%%%%%%%%%%%%%%%%%%%%%%%%%%%%%%%%%%%%%%%%%%%%%%%%%%%%%
%%%%%%%%%%%%%%%%%%%%%%%%%%%%%%%%%%%%%%%%%%%%%%%%%%%%%%%%%%%%%%%%%%%%%%%%%%%%%%%%%%%%%%%%%%%%%%%%%

\section{Discussion and conclusions}  \label{sec:concl}

We have shown that the unidimensional distance geometry problem %in one dimension 
can be cast as a problem of multiplying vectors and matrices. We have analyzed a classical optics computation approach that allows performing these multiplications in a massive parallel environment. We tool inspiration from the system presented in  \cite{feldmann2021}, but where we used the transverse spatial degrees of freedom of light {\it au lieu} of the longitudinal temporal ones. As a result, the approach becomes much simpler and technically much more accessible, while performing in comparable time scales.

Optical computation schemes for NP-hard problems (such as the one proposed and studied in this work) show that classical approaches exploring the wave properties of light can be helpful in order to access and realistically treat hard problems in large scale.

 \begin{acknowledgements}
Funding was provided by 
Coordena\c c\~{a}o de Aperfei\c coamento de Pessoal de N\'\i vel Superior (CAPES), 
Funda\c c\~{a}o de Amparo \`{a} Pesquisa do Estado de Santa Catarina (FAPESC),
Conselho Nacional de Desenvolvimento Cient\'{\i}fico e Tecnol\'ogico (CNPq), 
Instituto Nacional de Ci\^encia e Tecnologia de Informa\c c\~ao Qu\^antica (INCT-IQ 465469/2014-0),
and the ANR French funding agency (project {\sc multiBioStruct}, ANR-19-CE45-0019).
 \end{acknowledgements}

%\bibliographystyle{apsrev}

%%%%%%%%%%%%%%%%%%%%%%%%%%%%%%%%%%%%%%%%%%%%%%%%%%%%%%%%%%%%%%%%%%%%%%%%%%%%%%%%%%%%%%%%%%%%%%%%%
%%%%%%%%%%%%%%%%%%%%%%%%%%%%%%%%%%%%%%%%%%%%%%%%%%%%%%%%%%%%%%%%%%%%%%%%%%%%%%%%%%%%%%%%%%%%%%%%%
%%%%%%%%%%%%%%%%%%%%%%%%%%%%%%%%%%%%%%%%%%%%%%%%%%%%%%%%%%%%%%%%%%%%%%%%%%%%%%%%%%%%%%%%%%%%%%%%%

\appendix
\section*{Appendix~A}

We will give in the following a detailed example of a DGP$_1$ paradoxical instance and of the use of the BP algorithm for its solution. The following drawing shows the expected solution to our paradoxical DGP$_1$ example.
\begin{center}
\begin{tikzpicture}[scale=0.8]
\draw[line width=0.1pt,color=black] (0,0) -- (10,0);
\filldraw[color=gray] (5,0) circle (2pt);
\filldraw[color=blue] (1,0) circle (2pt);
\filldraw[color=brown] (3,0) circle (2pt);
\filldraw[color=red] (6,0) circle (2pt);
\end{tikzpicture}
\end{center}
We suppose that the vertices are sorted by considering the following vertex order: gray $\longrightarrow$ blue $\longrightarrow$ brown $\longrightarrow$ red. We also suppose that the distance between every vertex and its preceding vertex (except the gray vertex) is available, as well as the distance between the gray and the red vertex (in other words, this instance satisfies the assumptions in Def.~\ref{def:paradoxDGP1}).

In order to find possible solutions to our instance, the BP algorithm initially places at the center of the coordinate system (in the position of coordinate~0) the first vertex, and then it generates the candidate positions for the subsequent vertices by exploiting the available distance information. For example, there exist two possible positions for the blue vertex, and, for every selected position of the blue vertex, there are other two positions for the brown vertex, and so on.
\begin{center}
\begin{tikzpicture}[scale=0.45]
\draw[line width=0.2,color=gray] (5,0) -- (1,-1);  // distance is 4
\draw[line width=0.2,color=gray] (5,0) -- (9,-1);
\filldraw[color=gray] (5,0) circle (2pt);
\draw[line width=0.2,color=gray] (1,-1) -- (-1,-2);  // distance is 2
\draw[line width=0.2,color=gray] (1,-1) -- (3,-2);
\draw[line width=0.2,color=gray] (9,-1) -- (7,-2);
\draw[line width=0.2,color=gray] (9,-1) -- (11,-2);
\filldraw[color=blue] (1,-1) circle (2pt);
\filldraw[color=blue] (9,-1) circle (2pt);
\draw[line width=0.2,color=gray] (-1,-2) -- (-4,-3);  // distance is 3
\draw[line width=0.2,color=gray] (-1,-2) -- (2,-3);
\draw[line width=0.2,color=gray] (3,-2) -- (0,-3);
\draw[line width=0.2,color=gray] (3,-2) -- (6,-3);
\draw[line width=0.2,color=gray] (7,-2) -- (4,-3);
\draw[line width=0.2,color=gray] (7,-2) -- (10,-3);
\draw[line width=0.2,color=gray] (11,-2) -- (8,-3);
\draw[line width=0.2,color=gray] (11,-2) -- (14,-3);
\filldraw[color=brown] (-1,-2) circle (2pt);
\filldraw[color=brown] (3,-2) circle (2pt);
\filldraw[color=brown] (7,-2) circle (2pt);
\filldraw[color=brown] (11,-2) circle (2pt);
\filldraw[color=red] (-4,-3) circle (2pt);
\filldraw[color=red] (2,-3) circle (2pt);
\filldraw[color=red] (0,-3) circle (2pt);
\filldraw[color=red] (6,-3) circle (2pt);
\filldraw[color=red] (4,-3) circle (2pt);
\filldraw[color=red] (10,-3) circle (2pt);
\filldraw[color=red] (8,-3) circle (2pt);
\filldraw[color=red] (14,-3) circle (2pt);
\end{tikzpicture}
\end{center}
The tree above shows all the possible positions, for every vertex in the DGP instance, that can be constructed by using the entire distance information except the distance between the gray (the first) and the red (the last) vertex. The fact that our instance is in dimension~1 allows us to represent this tree with a drawing where the distances between possible positions for the same vertex are preserved in all parallel axes passing through the vertex positions (it is supposed that the axis passing through the only position for the gray vertex is parallel to all others). In this way, the {\it entire} set of possible positions for all the vertices in the instance can be obtained by projecting all these positions on one unique parallel line:
\begin{center}
\begin{tikzpicture}[scale=0.45]
\draw[line width=0.1pt,color=black] (-4,0) -- (14,0);
\filldraw[color=gray] (5,0) circle (3pt);
\filldraw[color=blue] (1,0) circle (3pt);
\filldraw[color=blue] (9,0) circle (2pt);
\filldraw[color=brown] (-1,0) circle (2pt);
\filldraw[color=brown] (3,0) circle (3pt);
\filldraw[color=brown] (7,0) circle (2pt);
\filldraw[color=brown] (11,0) circle (2pt);
\filldraw[color=red] (-4,0) circle (2pt);
\filldraw[color=red] (2,0) circle (2pt);
\filldraw[color=red] (0,0) circle (2pt);
\filldraw[color=red] (6,0) circle (3pt);
\filldraw[color=red] (4,0) circle (2pt);
\filldraw[color=red] (10,0) circle (2pt);
\filldraw[color=red] (8,0) circle (2pt);
\filldraw[color=red] (14,0) circle (2pt);
\end{tikzpicture}
\end{center}
Notice that we have marked with a larger circle the positions belonging to the expected solution represented above. The interesting question is: how to select our solutions from the set of given solutions? Every branch of the tree above gives in fact a set of possible positions for the 4 vertices in our instance that satisfies all distances that we have considered (up to now).

The selection of the actual solution set can be in fact performed by exploiting the distance between the gray and the red vertex (which has not been used yet). As it is possible to remark from the drawing above depicting one expected solution, the distance between the gray and red vertex is rather small, so that only two branches of the tree are likely to satisfy this distance constraint. The final solution set therefore contains two solutions, the expected one, plus its symmetric w.r.t the position of the first vertex:
\begin{center}
\begin{tikzpicture}[scale=0.80]
\draw[line width=0.1pt,color=black] (0,0) -- (10,0);  // the expected solution
\filldraw[color=gray] (5,0) circle (2pt);
\filldraw[color=blue] (1,0) circle (2pt);
\filldraw[color=brown] (3,0) circle (2pt);
\filldraw[color=red] (6,0) circle (2pt);
\draw[line width=0.1pt,color=black] (0,-1) -- (10,-1);  // and its symmetric
\filldraw[color=gray] (5,-1) circle (2pt);
\filldraw[color=blue] (9,-1) circle (2pt);
\filldraw[color=brown] (7,-1) circle (2pt);
\filldraw[color=red] (4,-1) circle (2pt);
\end{tikzpicture}
\end{center}
As already remarked in the Introduction (and this is the reason why we named the instances defined in Def.~\ref{def:paradoxDGP1} the {\it paradoxical instances}), the instance considered in our example has a particular difficult structure. In fact, while the final solution set only contains two symmetric solutions, the number of candidate solutions, only one layer before reaching the tree leaf nodes, is $2^{n-1}$. In our small example this exponential number is rather small, but it can become huge with a little larger instances. With the introduction of the fictive vertex ``$n+1$'' (see Section~\ref{sub:reformulation}), we can reduce the pruning phase of the algorithm to an additional branching phase, with a verification on the position of this fictive vertex. This brings to a total complexity of $2^n$, that our optic scheme can ideally lower down to $1$ when the matrix-by-vector multiplication can perform all operations detailed in this Appendix with a unique light.

\end{document}